\documentclass[aps,twocolumn]{revtex4-1}

\usepackage{bm}
\usepackage{amsmath}
\usepackage{upgreek}
\usepackage{graphicx}
\usepackage[subrefformat=parens,labelformat=parens]{subfig}
\usepackage{xcolor}
\usepackage{hyperref}

\usepackage{setspace}
\usepackage{footmisc}
\usepackage[makeroom]{cancel}
\usepackage{comment}
\def\be{\begin{equation}}
\def\ee{\end{equation}}
\def\ba{\begin{eqnarray}}
\def\ea{\end{eqnarray}}

\usepackage{graphicx}
\usepackage{amsmath,bbm}
\usepackage{amssymb,bm}

\begin{document}

\title{Anisotropic flow and correlations between azimuthal anisotropy Fourier harmonics in Xe-Xe collisions at $\sqrt{s_{NN}}$ = 5.44 TeV under HYDJET++ framework}
\date{\vspace{-15ex}}

\author{Saraswati Pandey$^{1}$\footnote{saraswati.pandey13@bhu.ac.in} and B. K. Singh$^{1}$\footnote{bksingh@bhu.ac.in}}

\affiliation{$^{1}$Department of Physics, Institute of Science, Banaras Hindu University,Varanasi, 221005, INDIA} 

\begin{abstract}
\noindent
The study of anisotropic harmonic flow coefficients $ v_{n}$(n=2,3,4) is performed in Xe-Xe collisions at $\sqrt{s_{NN}}$ = 5.44 TeV under Monte Carlo HYDJET++ model (HYDrodynamics plus JETs) framework. Anisotropic flow of identified particles and correlation between the azimuthal harmonic flow amplitudes is presented. Here, we have considered body-body and tip-tip type of geometrical configurations for Xe-Xe collision systems. The kinematic ranges $|\eta|<0.8$, $0<p_{T}<5.0$ GeV/c, and $|\delta / \eta|> 2$ are considered. The results have been shown for seven classes of centrality and compared with the ALICE experimental data. The anisotropic flow of identified charged particles show a strong centrality dependence. Mass ordering is observed for $v_{2},v_{3}$ and $v_{4}$. Mass ordering is different for different ranges of transverse momentum $p_{T}$. Strong correlation is observed between $v_{3}-v_{2}$, $v_{4}-v_{2}$, and $v_{4}-v_{3}$. Such correlation is centrality dependent and is different in different centrality windows. The anisotropic flow coefficients show a clear dependence on the total charged particle multiplicity. HYDJET++ model justifies experimental data well enough.

\end{abstract}

\date{\today}
\maketitle 

\section{Introduction}
\label{intro}
Several fascinating and impeccable phenomenon, yet not studied systematically, have been observed in the Relativistic Heavy Ion Collider (RHIC) and Large Hadron Collider (LHC) heavy-ion program. Experiments involved in the investigation of ultra-relativistic collisions aim to explore a deconfined state of quarks and gluons called as the Quark Gluon Plasma (QGP) \cite{doi:10.1146/annurev-nucl-101917-020852}. Quark Gluon Plasma is a new state of nuclear matter existing at high temperatures and densities, formed when the composite states of matter (hadrons) lose their identity and dissolve into a soup of quarks and gluons \cite{Aidala2019}. This  Quark Gluon Plasma (QGP) created in high energy heavy ion collisions expands rapidly. Relativistic viscous hydrodynamic models \cite{GALE_2013, Heinz_2013, Jia_2014} elegantly describe the space-time dynamics of this created Quark Gluon Plasma. During expansion, large pressure gradients of the generated QGP convert the spatial anisotropies in the initial-state geometry to the momentum anisotropies of the final state particles. These momemtum anisotropies are characterized by the Fourier expansion of particle density in the azimuthal angle $\phi$ \cite{voloshin1996flow, 2010rhip.book.....S},

\begin{equation}
\dfrac{dN}{d\phi} \propto 1+ 2\sum\limits_{n=1}^{\infty} v_{n}\cos[n(\phi-\psi_{n})] \quad
\end{equation}

where $\phi$ = azimuthal angle with respect to the reaction plane $\psi_{n}$ of the produced particle,\\
n = harmonic value, \\
$\psi_{n}$ = reaction plane, and\\
$v_{n}$ = fourier coefficient of order n representing the flow harmonics given by-
\begin{equation}
v_{n} = \langle\langle \cos[n(\phi-\psi_{n})] \rangle\rangle \quad.
\end{equation}

The second harmonic, n=2 is called as elliptic flow $v_{2}$ that reveals the lenticular shape of the collision overlap region. The elliptic flow relates the anisotropic shape of the overlapped region of the colliding nuclei to the corresponding anisotropy of the outgoing momentum distribution. The higher harmonics $v_{n}$ (n$>$2) are produced lesser than $v_{2}$. These coefficients also carry essential information on the dynamics of the created medium and provide a more clear and complete picture of its bulk properties along with $v_{2}$. The third harmonic, n=3 is called as triangular flow $v_{3}$ and the fourth harmonic, n=4 is called as quadrangular flow $v_{4}$. Triangular flow is caused due to the initial-state fluctuations in the nucleon positions at the moment of impact \cite{PhysRevC.81.054905} while the higher harmonics are affected by the dynamics of the expanding system. These harmonics have been closely studied so far \cite{PhysRevC.68.031902, PhysRevC.69.051901, PhysRevC.81.054905}. The pentagonal and hexagonal flows $v_{5}$ and $v_{6}$ respectively are studied to a lesser extent may be due to lack of experimental evidences or so, some predictions from hydrodynamics on them also have been made \cite{PhysRevC.82.034913}. At relatively low $p_{T}$, pressure driven anisotropic expansion of the created matter results in the azimuthal anisotropy, emitting more particles in the direction of large pressure gradients \cite{PhysRevD.46.229}. At higher transverse momentum $p_{T}$, the anisotropy is explained using the path-length dependent energy loss of partonic jets as they traverse the matter, emitting more jet particles in the direction of shortest path-length \cite{PhysRevLett.86.2537}. Anisotropic flow develops in the system because the spatial anisotropies $\epsilon_{n}$ of the overlapping region are transformed into the momentum anisotropies $v_{n}$ of final hadronic distribution. Due to the initial state fluctuations, these spatial anisotropies exist even in very central collisions. Momentum anisotropy also arises from the non-isotropic azimuthal dependence of the transverse velocity of the expanding fireball resulting in collective flow gradients in various directions. These two different sources of the particle momentum anisotropy are called as geometric and dynamical anisotropy, respectively. It might be possible to disentengle these anisotropies by the simultaneous analysis of the flow harmonics, which we aim to perform in our present work.

Recently, we performed our study on Xe-Xe collisions, where we emphasized on the motto of performing our analysis on xenon-xenon collision systems \cite{PhysRevC.103.014903}. The similarities observed between smaller systems, such as p + p and p + Pb and larger Pb-ion systems are debatable as to whether they arise from same physics mechanism. Here, again we will deal with body-body and tip-tip collision configurations \cite{PhysRevC.85.034905}. The charged particle multiplicity density in the transverse phase space is higher in deformed collision systems than spherical or non-deformed nucleus collisions \cite{PhysRevC.73.034911, TRIPATHY201881}. Lately, HYDJET++ model was modified to study U-U collisions at 193 GeV center-of-mass energy in body-body and tip-tip geometrical configurations \cite{singh2018transverse}. Anisotropic flow harmonics and the correlations between them have been studied intensively at both RHIC and LHC energies. Recently, \cite{PhysRevC.97.034904, GIACALONE2019371}, predictions were made for Xe-Xe collisions using event-by-event hydrodynamic simulations discussing anisotropic flow coefficients as a function of centrality and comparing the results with the ATLAS, CMS, ALICE experimental measurements. From the results of anisotropic flow coefficients for Xe-Xe and Pb-Pb collisions, it was confirmed that xenon has a non-spherical shape. In references \cite{PhysRevC.102.044905, SCHENKE2021121756, PhysRevC.103.054902}, anisotropic flow coefficients studies were performed in O-O, Al-Al, and Cu-Cu collisions under Color Gluon Condensate (CGC) framework or under AMPT model (a multiphase transport model) approach or in the fusing color string model. Here, a thorough description of bulk observables and multi-particle correlations in several collision systems such as Au+Au, U+U, Ru+Ru, Zr+Zr, and O+O collisions at top RHIC energies and Pb+Pb, Xe+Xe, and O+O collisions at LHC energies are performed. In reference \cite{PhysRevC.99.024903}, measurements of elliptic and triangular azimuthal anisotropy of charged particles in Au+Au collisions were presented using the multiparticle cumulant technique to study the centrality dependence of $v_{2}$ and $v_{3}$ along with the significances of initial geometrical fluctuations and their translation into the final state momentum distributions. An investigation was performed in Xe-Xe collision systems on the system-size dependence of the longitudinal decorrelations of $v_{2},v_{3}$ and $v_{4}$ and comparing the results with Pb-Pb collisions at 5.02 TeV \cite{PhysRevLett.126.122301}. Recently, measurements of anisotropic flow harmonic coefficients ($v_{n}$) for inclusive charged particles and identified hadrons were performed in Cu+Au (asymmetric) collisions at 200 GeV \cite{PhysRevC.94.054910} where mass ordering in hydrodynamic flow and the particle azimuthal distributions were studied as a function of $p_{T}$ over various centrality classes at RHIC energies. In another work, mass dependency of hadrons (pions, kaons and protons) for elliptic flow $v_{2}$($p_{T}$) is observed similar to the observations from Pb-Pb collisions \cite{Tripathy_2018}. Such study has not been performed for the higher flow harmonics. Correlation between anisotropic flow coefficients $v_{n}$ and average transverse momentum $\langle p_{T} \rangle$ of outgoing particles in Pb + Pb collisions is studied at LHC energy tracing back to the initial density profile, i.e., to the early stages of the collision \citep{PhysRevC.103.024909}. A lot of study in the above mentioned aspect has been done to understand higher flow harmonics in spherical collision systems at RHIC as well as LHC energies but deformed systems have not been touched much till now, especially visualisation in various geometrical configurations.  

Azimuthal correlations provide valuable information about the relativistic hydrodynamic nature of the medium, about its transport coefficients, and also about the fluctuations in the initial state from which the medium is formed \cite{braun2016properties}. They are extensively studied as a function of centrality of collision and transverse momentum $p_{T}$ \cite{PhysRevC.100.044902, PhysRevC.103.034905}, produced particle type, rapidity, and expected event-by-event geometrical fluctuations of the nuclei \cite{doi:10.1146/annurev-nucl-102212-170540}. It is to be noted that \cite{PhysRevC.68.031902, BORGHINI2006227} in the absence of event-by-event fluctuation (involving hydrodynamics with smooth initial condition), the even flow harmonic coefficients are found to be correlated. The reason being that despite the fluid velocity profile is elliptically deformed, complete set of the even flow harmonic coefficients is generated in general as fluid velocity enters as the exponent of the (flow-boosted) thermal distribution on the freeze-out surface. When these event-by-event fluctuations in the initial state are present, the resulting flow fluctuations of different harmonic orders are generally correlated by the geometric constraints on the shapes and positions of these fluctuations within the overlapping spatially deformed region. The availability of experimental data from various collision programmes gives a motivation to perform a study whether the experimentally measured anisotropic flow correlations and their dependence on centrality of collision can be understood and described well through a successful hydrodynamical model approach. In our study of disentangling both geometrical and dynamical anisotropies and performing analysis of the higher anisotropic flow coefficients we choose HYDJET++ Model framework \cite{lokhtin2009heavy}. This model allows to switch on/off both the anisotropy parameters independently. The production of higher flow harmonics within HYDJET++ framework is advantageous in the sense that the interplay of ideal hydrodynamics with jets reveal the role of hard processes in the production of secondary hadrons. Also, the existence of $v_{2}$ and $v_{3}$ allows us to understand the contribution of these to all other higher odd and even anisotropic flow coefficients \cite{PhysRevC.89.024909}. In reference \cite{Lokhtin_2012}, HYDJET++ model was used to study the LHC data on multiplicity, charged hadron spectra, elliptic flow and femtoscopic correlations in Pb-Pb collisions. Considering both soft as well as hard components along with the tuning of the parameters, we can reproduce the experimental data under HYDJET++. In this paper, we have studied the centrality, transverse momentum and total charged particle multiplicity dependence of anisotropic flow coefficients in Xe-Xe collisions at 5.44 TeV center-of-mass energy. $m_{T}$ dependence of anisotropic flow harmonics is visualized here. The correlation between these coefficients is an interesting part of this work. The analysis of our results have been performed in body-body and tip-tip geometrical configurations using the HYDJET++ model. In Sec. II, we have briefly discussed formulation of the HYDJET++ model and the incorporation of deformation in the body of the model. Also, we discuss how the model incorporates the higher flow harmonics in the body of the model. In Sec. III, we present the results and discussions part for the elliptic flow $v_{2}$, triangular flow $v_{3}$ and quadrangular flow $v_{4}$ distributions. Lastly, we have summarized our results in Sec. IV.

\section{Model Formalism}
\label{model}
\vspace{-2ex}
HYDJET++ (HYDrodynamics plus JETs) is a Monte Carlo model of relativistic heavy ion collisions which includes the simultaneous superposition of two independent components: the soft hydro-type state and the hard state resulting from the medium-modified multiparton fragmentation. The details of the model and the corresponding simulation procedure can be found in the paper \cite{lokhtin2009heavy, bravina2017dynamical} and the references there within. The model parameters have been tuned to reproduce the experimental LHC data on various physical observables measured in Xe-Xe collisions at 5.44 TeV of center-of-mass energy per nucleon pair. A concise view about the physics of the model valuable for our study has been presented in our previous article  \citep{PhysRevC.103.014903} where we have discussed about both hard as well as soft part of the model. In there, we performed elliptic flow studies showing the transverse momentum ($p_{T}$) and centrality dependence of elliptic flow $v_{2}$ but did not work on higher anisotropic fourier harmonics.

To simulate higher azimuthal anisotropy harmonics, various alterations were required and have been made in HYDJET++. Basically, the model does not involve the evolution of fireball from the initial state to the final state freeze-out stage. It utilizes simple and often used parameterization of the freeze-out hypersurface rather than using computational relativistic hydrodynamics (time consuming). The anisotropic elliptic shape of the initial overlap of the colliding nuclei results in a corresponding anisotropy of the outgoing momentum distribution. The second harmonic $v_{2}$ is described using the coefficients $\epsilon_{2}$(b) and $\delta_{2}$(b) known as the spatial anisotropy and momentum anisotropy, respectively. $\epsilon_{2}$(b) exemplifies the elliptic modulation of the final freeze-out hypersurface at a given impact parameter b, whereas $\delta_{2}$(b) deals with the alteration of flow velocity profile. These two parameters can be treated independently for each centrality or can be made interdependent via the dependence on the initial ellipticity $\epsilon_{0}(b)=b/2R_{A}$ where $R_{A}$ is the nucleus radius. Here, we are treating them independent of each other. The transverse radius of the fireball is given as:

\begin{equation}
R_{ell}(b,\phi)=R_{f}(b)\frac{\sqrt{1-\epsilon_{2}^{2}(b)}}{1+\epsilon_{2}(b)\cos 2\phi},
\end{equation} 
          
where,
\begin{equation}
R_{f}(b)=R_{0}\sqrt{1-\epsilon_{2}(b)}.
\end{equation}
           
Here $R_{0}$ denotes is the freeze-out transverse radius in absolute central collision with b=0. Then, the spatial anisotropy gets transformed into momentum anisotropy at freeze-out, because each of the fluid cells is carrying some momentum. The term dynamical anisotropy arises here implying that the azimuthal angle of the fluid cell velocity, $\phi_{cell}$ does not coincide with the azimuthal angle $\phi$, instead correlates with it \cite{amelin2008fast} through the non-linear function involving the anisotropy parameter $\delta_{2}(b)$

\begin{equation}
\frac{\tan \phi_{cell}}{\tan \phi}=\sqrt{\frac{1-\delta_{2}(b)}{1+\delta_{2}(b)}}.
\end{equation}  

In case where $\delta \neq 0$ even the spherically symmetric source can mirror the spatially contracted one. The elliptic flow coefficient $v_{2}(\epsilon, \delta_{2})$ in the hydrodynamical approach \citep{PhysRevC.57.266} is given as:

\begin{equation}
v_{2}(\epsilon_{2},\delta_{2}) \propto \frac{2(\delta_{2}-\epsilon_{2})}{(1-\delta_{2}^{2})(1-\epsilon_{2}^{2})}.
\end{equation}  

For triangular flow $v_{3}$ in HYDJET++, the model has another parameter $\epsilon_{3}$(b),for spatial triangularity of the fireball. Thus the modified radius of the freeze-out hypersurface in azimuthal plane reads:

\begin{equation}
R(b,\phi)=R_{ell}(b)\lbrace1+\epsilon_{3}(b)\cos[3(\phi-\psi_{3}^{RP})]+...\rbrace.
\end{equation}           
where,
$\phi$ = spatial azimuthal angle of the fluid element relatively to the direction of the impact parameter.

The phase $\psi_{3}^{RP}$ gives us the advantage to introduce a third harmonic having its own reaction plane, distributed randomly with respect to the direction of the impact parameter($\psi_{2}^{RP}=0$). This new anisotropy parameter, $\epsilon_{3}(b)$ again can be handled in two ways: independently for each centrality  and dependent using  $\epsilon_{0}(b)=b/2R_{A}$ where $R_{A}$ has its meaning unchanged. Such modifications do not affect the elliptic flow (controlled by $\epsilon(b)$ and $\delta(b)$). Hence, the triangular dynamical anisotropy can be incorporated by the parameterization of the maximal transverse flow rapidity [23],
\begin{equation}
\rho_{u}^{max}(b)=\rho_{u}^{max}(0)\lbrace 1+ \rho_{3u}(b)\cos [3(\phi-\psi_{3}^{RP})] +...\rbrace.
\end{equation}  

As a result, the maximal transverse flow rapidity \cite{lokhtin2009heavy} after the parameterization of the four-velocity $u$ upto the fourth order harmonics is given as,

\begin{equation}
\rho_{u}^{max}(b)=\rho_{u}^{max}(0)\lbrace 1+ \rho_{3u}(b)\cos 3\phi + \rho_{4u}(b)\cos 4\phi +...\rbrace.
\end{equation}  

Hence, we can calculate higher harmonics with respect to the direction of the impact parameter $b \psi_{2}^{RP}=0$. Again, these new anisotropy determiners $\rho_{3u}(b)$ and $\rho_{4u}(b)$ can be treated both independently and dependent via initial ellipticity $\epsilon_{0}(b)=b/2R_{A}$. Now, here we opted the former case and treated the parameters independently and varied them with centrality.

The next important part of the HYDJET++ model is the incorporation of the intrinsic deformation in Xe nucleus. This has been already done in our previous work \citep{PhysRevC.103.014903}, where we performed our study in both tip-tip and body-body geometrical configurations making our modified HYDJET++ model work at both RHIC as well as LHC energies. After implementing higher fourier harmonics the simulation and the optimization of the parameters is verified. We obtain results similar to the figures 2 and 3 in our previous work \cite{PhysRevC.103.014903} thereby certifying our HYDJET++ model simulations in both tip-tip and body-body type of geometrical configurations. 

\bigskip

\section{Results and Discussions}
\label{results}
\vspace{-2ex}

\begin{figure}[htbp]
\centering
\includegraphics[scale=0.3]{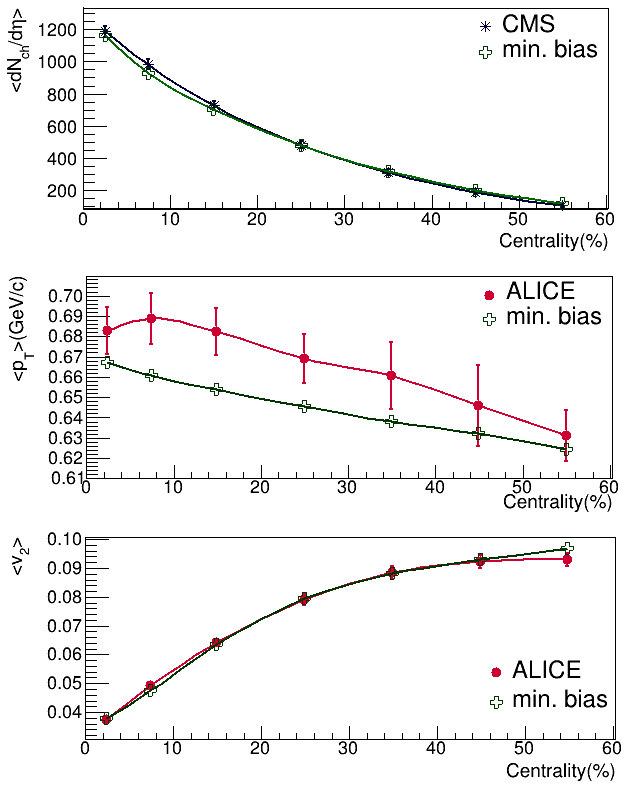} 
\caption{From article \citep{PhysRevC.103.014903}}
\label{fig:previous}
\end{figure}

We have generated $5 \times 10^5$ events using the modified HYDJET++ model in different centrality classes for both tip-tip and body-body configurations at 5.44 TeV center of mass energy. We have performed our simulations for n$\leq$4 as the model has been designed upto that only. Only those events have been considered for the results which fall in the kinematic range $|\eta|<0.8$ and $0<p_{T}<5$ GeV/c. In our previous work on Xe-Xe collision systems \cite{PhysRevC.103.014903}, it was demonstrated that tuned HYDJET++ model can reproduce the LHC data on centrality and transverse momentum  dependence of charged particle multiplicity density, transverse momentum $p_{T}$ spectra and elliptic flow coefficient $v_{2}$ up to $p_{T}\sim$2.0 GeV/c and 60\% centrality range). However, the reasonable treatment of higher Fourier harmonics of particle azimuthal distribution $v_{n}$ (n$>$2) needs additional modifications in the model, which does not affect our previous results. This is evident from figure \ref{fig:previous}. We have compared the results of HYDJET++ simulations with the LHC (ALICE) experimental data \cite{201882} on $v_{n} \lbrace2\rbrace$ second order cumulant for inclusive as well as for identified charged hadrons for our analysis.

\begin{figure}[htbp]
\centering
\includegraphics[scale=0.3]{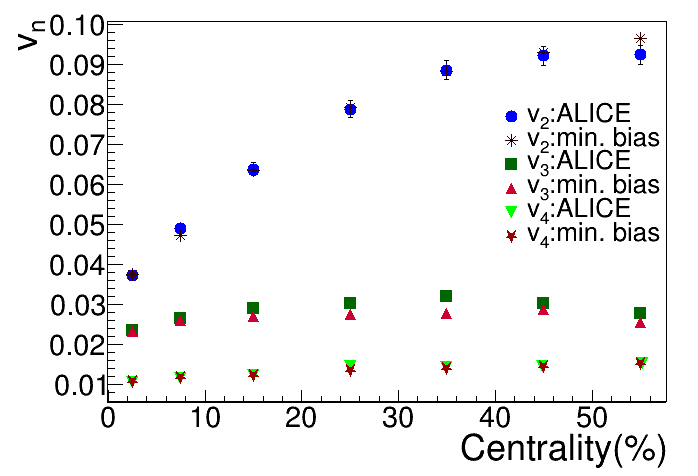} 
\caption{Centrality dependence of  $v_{n}$ along with ALICE experimental data \cite{201882}.}
\label{vn_bRA}
\end{figure}

Figure \ref{vn_bRA} presents the variation of azimuthal anisotropy fourier harmonics ($v_{2},v_{3}$ and $v_{4}$) in Xe-Xe collision systems at 5.44 TeV with centrality. The model results in minimum bias have been compared with the ALICE experimental data \cite{201882}. A strong centrality dependence of elliptic, triangular and quadrangular flows is observed here. There is a fair agreement of the HYDJET++ model results with the ALICE experimental data both qualitatively as well as quantitatively. As we move from most central to most peripheral collisions, elliptic flow increases and then decreases in most peripheral collisions. Similar behaviour is shown by triangular flow but the fall is seen here early (centrality $>$40\%). However, quadrangular flow shows a gradual increase as we move from most central to most peripheral class of collisions. The elliptic flow $v_{2}$ results from HYDJET++ match very well with the experimental data in all centrality windows except in most peripheral collisions where the model overpredicts the data. The triangular flow $v_{3}$ shows a good agreement with the experimental data in central collisions. As we move from most central, towards semi-peripheral collisions (centrality $<40\%$), the deviation from the experimental results is observed (model underpredicts the experimental data). However, the deviation decreases as we move towards most peripheral collisions (centrality $>40\%$). Lastly, as we move from most central to most peripheral collisions the quandrangular flow $v_{4}$ results from HYDJET++ show a suitable match with the ALICE experimental result at all centrality classes of Xe-Xe collisions.

\begin{figure}[htbp]
\centering
\includegraphics[scale=0.3]{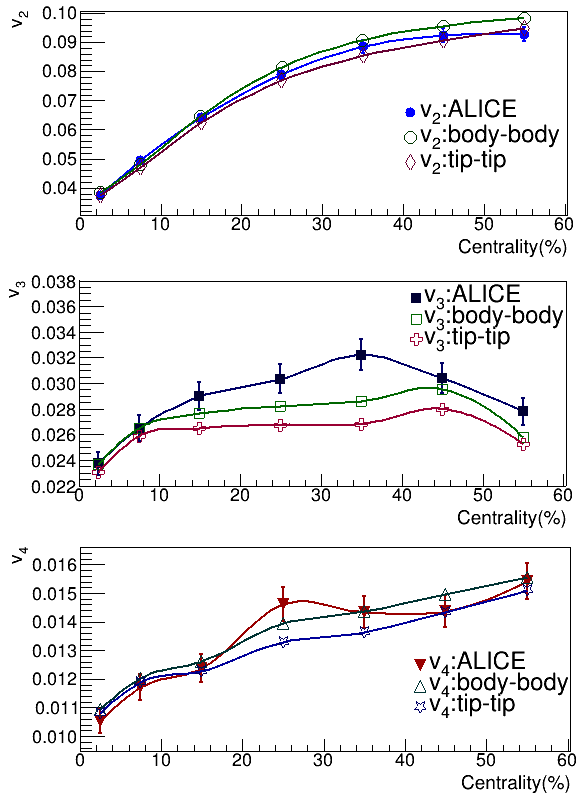} 
\caption{Centrality dependence of  $v_{n}$  for body-body and tip-tip geometrical configurations along with ALICE experimental data \cite{201882} for comparison.}
\label{vn_bRA_configs}
\end{figure}

In figure \ref{vn_bRA_configs}, we have compared our model results in body-body and tip-tip collisions with the ALICE experimental data for $v_{2},v_{3}$ and $v_{4}$. The anisotropic harmonic coefficients obtained from our HYDJET++ model in both the geometrical configurations show strong centrality dependence. The qualitative behaviour of the two geometrical configurations is similar to the the one presented in figure \ref{vn_bRA}. In case of elliptic flow $v_{2}$, we find that our results agree with experimental results both quantitatively as well as qualitatively. The body-body collision results are higher than tip-tip collision results. However, in most central collisions, there is hardly any difference between the two geometrical configurations. But this difference is clearly visible as we move towards peripheral collisions. In most peripheral collisions, model results in the two geometrical configurations are higher than data thereby overpredicting the experimental result. Moving to triangular flow $v_{3}$ results, our model results for the two geometrical configuartions match ALICE results qualitatively. However, in a closer view, we find that quantitatively HYDJET++ results for the two geometrical configurations underpredict the experimental data. Body-body collision results are higher than tip-tip collision results. The two geometrical configurations cannot be disentangled in most central and most peripheral collisions but the difference can be seen very clearly in semi-peripheral collisions. The HYDJET++ results for quadrangular flow $v_{4}$ in body-body and tip-tip collisions show a suitable match with the experimental data qualitatively. In central collisions (centrality $<20\%$), it is difficult to differentiate between the two geometrical configurations. This can be accomplished as we move towards peripheral collisions. A bump appears in between (20\%-30\%) class of collision centrality in the ALICE experimental data. This bump can be seen in our HYDJET++ model results, although not so prominent. Again, body-body collision results are higher than tip-tip colision results.

\begin{figure*}[htbp]
\centering
\includegraphics[scale=0.35]{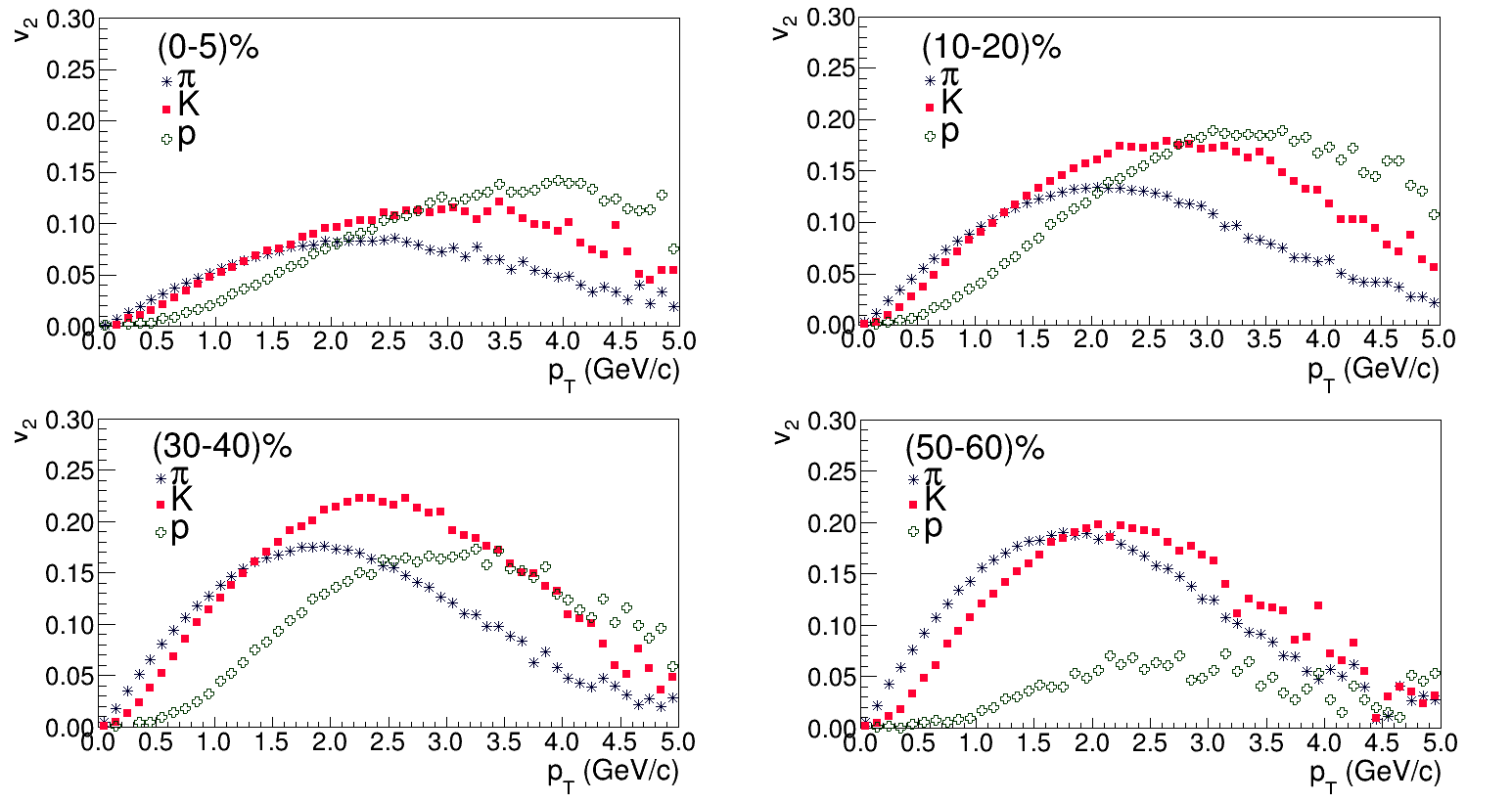} 
\caption{Transverse momentum dependence of  $v_{2}$ for identified particles in different centrality windows.}
\label{v2_pt}
\end{figure*}

Figure \ref{v2_pt} presents the elliptic flow $v_{2}$ of identified charged particles with respect to transverse momentum in four classes of centrality. As a function of centrality, elliptic flow $v_{2}$ increases as we move from most central collisions to semi-peripheral collisions and then starts to decrease as we enter region of most peripheral ((50-60)\%) class of collision. Mass ordering is observed here in each class of collision. At low $p_{T}$, ($p_{T}<$1.6 GeV/c) $v_{2}$ for the lower mass particle (pions) is more than the higher mass particles. In simple words, $v_{2}^{\Pi}>v_{2}^{K}>v_{2}^{p}$ where $m^{\Pi}<m^{K}<m^{p}$. However, for $p_{T}\geq$ 1.6 GeV/c the situation changes. As we move from most central to peripheral collisions, mass ordering reverses. Higher mass particles are produced more. $v_{2}^{\Pi}<v_{2}^{K}<v_{2}^{p}$ where $m^{\Pi}<m^{K}<m^{p}$. In most peripheral collisions, the scenario appears completely different. At $p_{T}>$1.6 GeV/c $v_{2}^{K}>v_{2}^{\Pi}>v_{2}^{p}$ where $m^{\Pi}<m^{K}<m^{p}$. This cut value of transverse momentum may vary from centrality to centrality. At much higher $p_{T}\geq 2.8\pm0.2$, $v_{2}^{p}>v_{2}^{K}>v_{2}^{\Pi}$ in most central and semi-peripheral class of collisions. This is absent in peripheral collisions and $v_{2}^{p}$ is found to be lesser.

\begin{figure*}[htbp]
\centering
\includegraphics[scale=0.35]{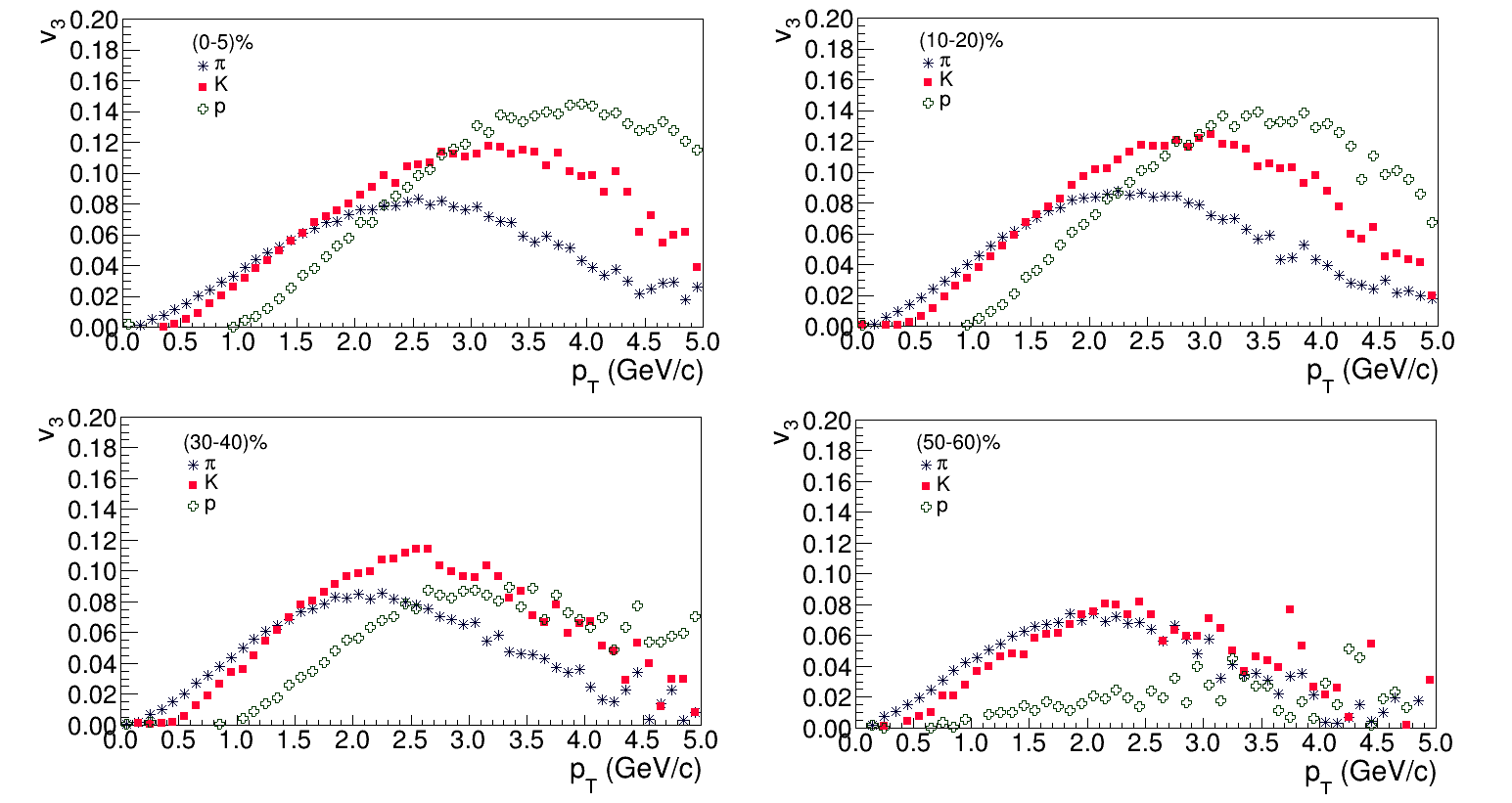} 
\caption{Transverse momentum dependence of  $v_{3}$ for identified particles in different centrality classes.}
\label{v3_pt}
\end{figure*}

Figure \ref{v3_pt} shows transverse momentum dependence of triangular flow $v_{3}$ for identified charged particles in four classes of centrality. Triangular flow $v_{3}$ increases as we move from most central collisions and decreases as we move towards most peripheral class of collisions. Mass ordering is observed here in each class of collision. At low $p_{T}$, ($p_{T}<$1.5 GeV/c) $v_{3}$ for the lower mass particle (pions) is more than $v_{3}$ for higher mass particles. In simple words, $v_{3}^{\Pi}>v_{3}^{K}>v_{3}^{p}$ where $m^{\Pi}<m^{K}<m^{p}$. However, for $p_{T}\geq$1.5 GeV/c the situation changes. As we move from most central to peripheral collisions, mass ordering reverses. Higher mass particles are produced more. $v_{3}^{\Pi}<v_{3}^{K}<v_{3}^{p}$ where $m^{\Pi}<m^{K}<m^{p}$. In most peripheral collisions, the scenario appears different. At $p_{T}>$1.7 GeV/c $v_{3}^{K}>v_{3}^{\Pi}>v_{3}^{p}$ where $m^{\Pi}<m^{K}<m^{p}$. At much higher values of $p_{T}\geq 2.8\pm0.2$, again $v_{2}^{p}>v_{2}^{K}>v_{2}^{\Pi}$ in most central and semi-peripheral class of collisions, and in peripheral collisions $v_{2}^{p}$ is found to be lesser.

\begin{figure*}[htbp]
\centering
\includegraphics[scale=0.35]{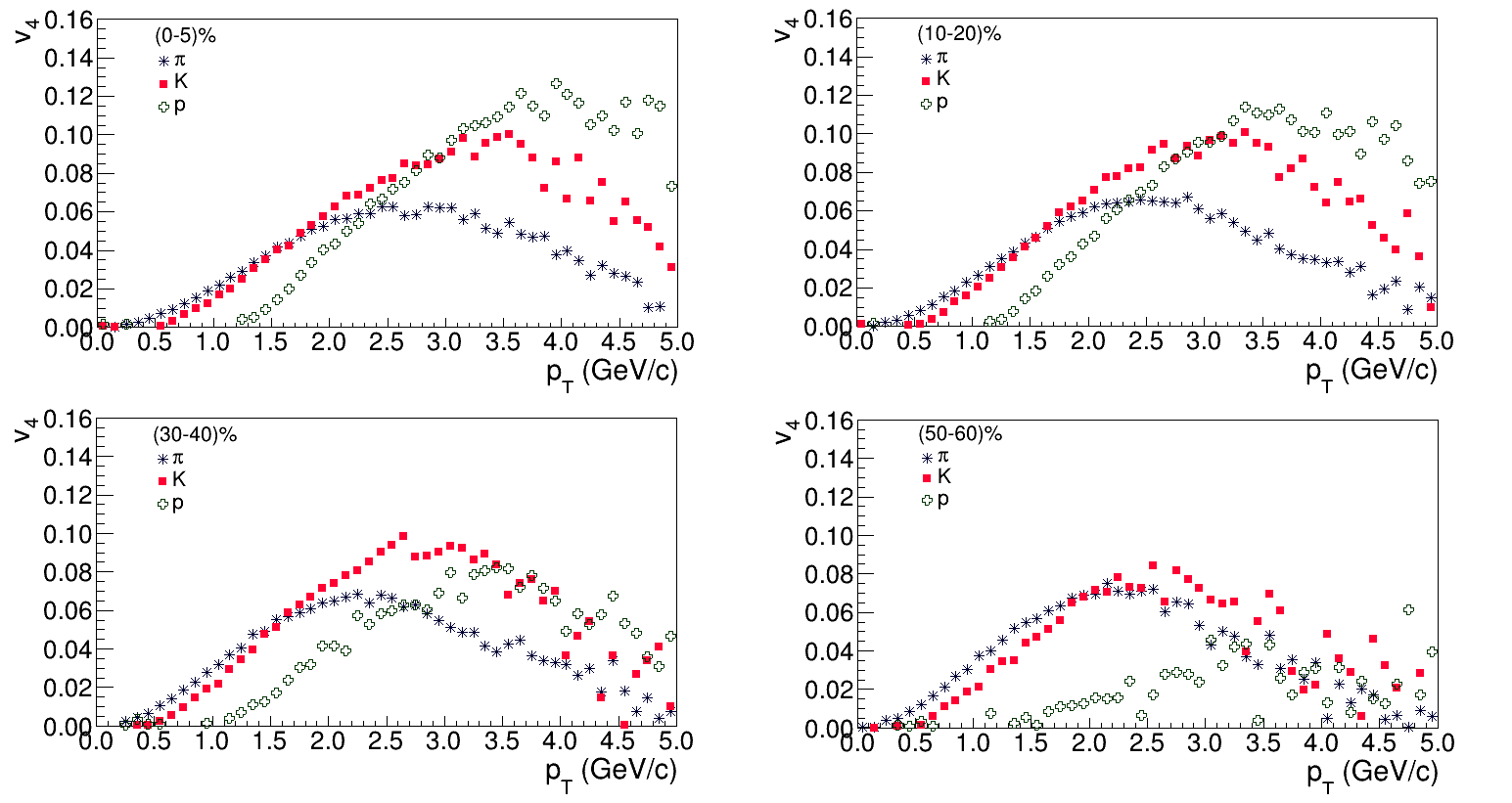} 
\caption{Transverse momentum dependence of  $v_{4}$ for identified particles in different centrality windows.}
\label{v4_pt}
\end{figure*}

Figure \ref{v4_pt} presents the quadrangular flow $v_{4}$ of identified charged particles with respect to transverse momentum in various centrality windows. Quadrangular flow $v_{4}$ increases as we move from most central collisions and decreases as we move towards most peripheral class of collisions. Mass ordering is observed here in each class of collision. At low $p_{T}$, ($p_{T}<$1.7 GeV/c) $v_{4}$ for the lower mass particle (pions) is more than $v_{4}$ for higher mass particles. In simple words, $v_{4}^{\Pi}>v_{4}^{K}>v_{4}^{p}$ where $m^{\Pi}<m^{K}<m^{p}$. However, for $p_{T}>$1.7 GeV/c the situation changes. As we move from most central to peripheral collisions, mass ordering reverses. Higher mass particles are produced more. $v_{4}^{\Pi}<v_{4}^{K}<v_{4}^{p}$ where $m^{\Pi}<m^{K}<m^{p}$. In most peripheral collisions, the scenario appears different. At $p_{T}>$1.7 GeV/c $v_{4}^{K}>v_{4}^{\Pi}>v_{4}^{p}$ where $m^{\Pi}<m^{K}<m^{p}$. Similar to the observations in figures \ref{v2_pt} and \ref{v3_pt}, here too we have $v_{2}^{p}>v_{2}^{K}>v_{2}^{\Pi}$ in most central and semi-peripheral collisions at $p_{T}\geq 2.8\pm0.2$, this being absent in peripheral class of collisions. 

Thus, from the above anaysis of the results in figures \ref{v2_pt}, \ref{v3_pt}, and \ref{v4_pt} we conclude that the $p_{T}$ cut value is different for different flow coefficients but the qualitative behaviour of the identified particles and mass ordering of $v_{n}$ for n=2,3,4 is similar.

\begin{figure}[htbp]
\centering
\includegraphics[scale=0.3]{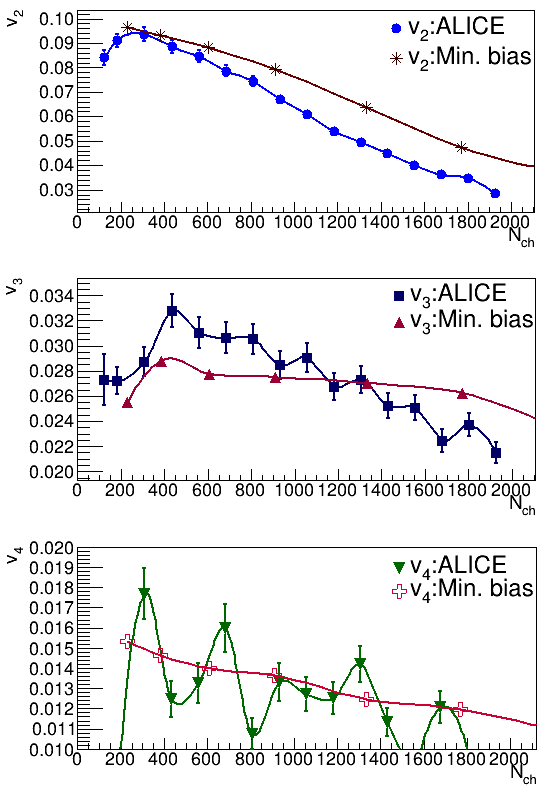} 
\caption{Variation of  minimum bias $v_{n}$  with total charged particle multiplicity along with ALICE experimental data \cite{201882} for comparison.}
\label{vn_nch}
\end{figure}

In figure \ref{vn_nch}, we present the variation of minimum bias $v_{n}$  with total charged particle multiplicity. Here we have compared our HYDJET++ model results with the ALICE experimental data \cite{201882}. The qualitative behaviour for different flow coefficients, $v_{2},v_{3}$ and $v_{4}$ is similar to ALICE experimental results. The elliptic flow $v_{2}$ decreases gradually as the total charged particle multiplicity increases. In most peripheral class of collisions, the model results are in very close agreement with the experimental data. As we move towards most central class of collisions, the observed deviation from the experiment results increases. The triangular flow $v_{3}$ shows a linear increase and then falls gently as we move to higher values of total charged particle multiplicity. A soft peak is observed at $n_{ch}\approx400$. Quantitatively, our HYDJET++ results underestimate the ALICE experimental data in all classes of collisions except in most central collisions where our HYDJET++ model results overestimate the experimental results. The plot of quadrangular flow $v_{4}$ with respect to the total charged particle multiplicity shows that the $v_{4}$ decreases as we move from most peripheral to most central class of collisions. However, due to lack of results in some more centrality classes, we do not obtain the exactly similar behaviour. But we can say that our HYDJET++ results match the experimental results quantitatively with least errors.

\begin{figure}[htbp]
\centering
\includegraphics[scale=0.3]{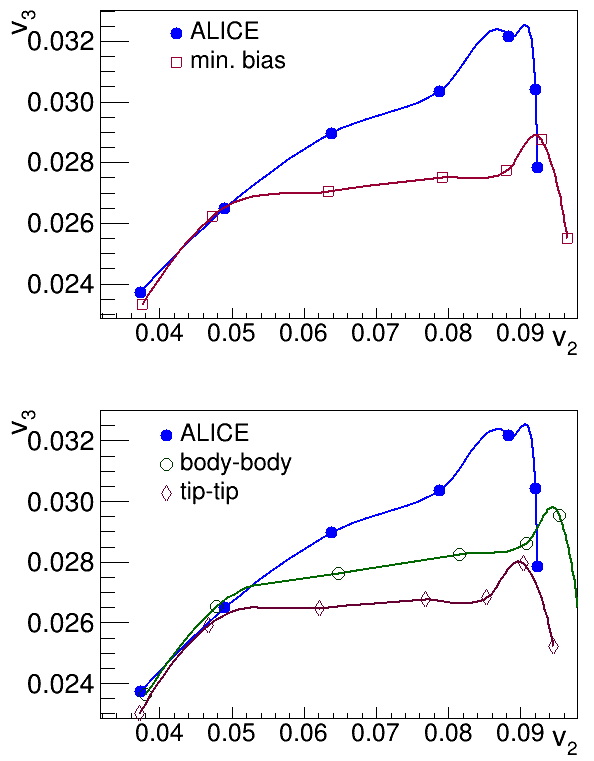} 
\caption{Correlation between $v_{3}$ and $v_{2}$ for $p_{T}<$2 GeV/c over seven centrality classes in Xe-Xe collisions at 5.44 TeV. The results have been compared with ALICE experimental data \cite{201882}.}
\label{v3_v2}
\end{figure}

Figure \ref{v3_v2} presents the correlation between $v_{3}$ and $v_{2}$ for seven centrality windows from our HYDJET++ model and from experimental data \cite{201882} for Xe-Xe collisions at the LHC. Here, we observe a qualitative agreement of our model results with the experimental data. The upper panel of the figure shows the comparison of minimum bias results with the experimental data. In central collisions, a linear positive correlation is observed between $v_{3}$ and $v_{2}$. Also, our model results show suitable match with the experimental data quantitatively. However, in mid-central or semi-peripheral collisions, HYDJET++ results underpredict the data quantitatively. The correlation between $v_{3}$ and $v_{2}$ is not very much positive due to the reason that our model fails to predict $v_{3}$ in these collision centralities. As we move towards peripheral collisions, again positive correlation is seen between $v_{3}$ and $v_{2}$. In most peripheral collisions, the situation changes, a sharp negative correlation is predicted which is in good agreement with the ALICE experimental results qualitatively. Quantitatively, there exists a clear deviation of our model results from the experimental results in most peripheral collisions. This is because HYDJET++ model fails to handle such collision centralities. Hence, the correlation structure between $v_{3}$ and $v_{2}$ as a function of centrality might be attributed to the fact that $v_{3}$ has weaker centrality dependence as compared to $v_{2}$.

The lower panel of figure \ref{v3_v2} shows the correlation between $v_{3}$ and $v_{2}$ for seven classes of centrality from our HYDJET++ model in body-body and tip-tip geometrical configurations. These results have been compared with the ALICE experimental data \cite{201882} where a complete agreement is observed between model and experiment qualitatively. Quantitatively, body-body results are higher than tip-tip results. In central collisions, the correlation between $v_{3}$ and $v_{2}$ in both the cases is similar to the above described for the upper panel results. However, in mid-central or semi-peripheral collisions, body-body and tip-tip collision results show a positive correlation between $v_{3}$ and $v_{2}$, tip-tip being weaker than body-body collisions. This correlation indicates a similar correlation between the initial eccentricities $\epsilon_{3}$ and $\epsilon_{2}$ \cite{PhysRevC.92.034903, PhysRevC.90.024910}. Such inference is expected because the hydrodynamic response of $v_{3}$ and $v_{2}$ to $\epsilon_{3}$ and $\epsilon_{2}$ , respectively, is linear approximately, especially at small eccentricities. A small dip is observed in the ALICE experimental result as the correlation changes from positive to negative (as we move from peripheral to most-peripheral collisions). This dip is not so prominent in minimum bias and body-body collision results but can be seen in tip-tip collisions.

\begin{figure}[htbp]
\centering
\includegraphics[scale=0.3]{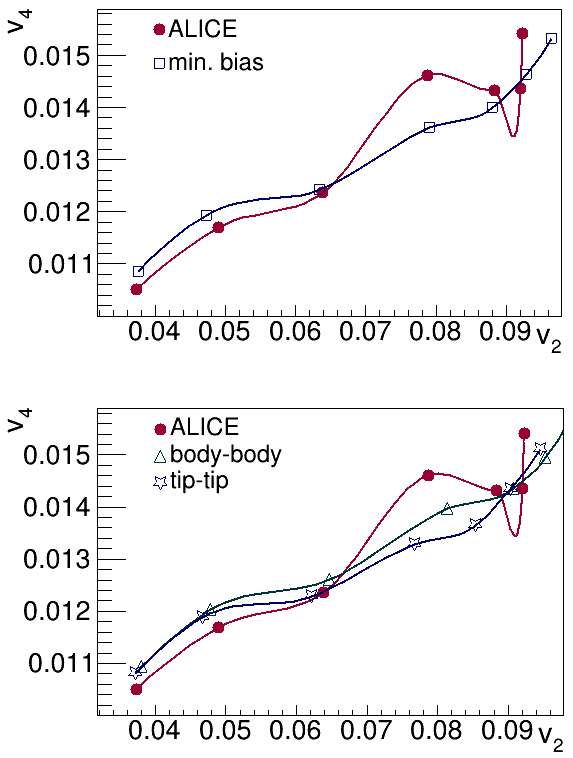} 
\caption{Correlation between $v_{4}$ and $v_{2}$ for $p_{T}<$2 GeV/c over seven centrality classes in Xe-Xe collisions at 5.44 TeV. The results have been compared with ALICE experimental data \cite{201882}.}
\label{v4_v2}
\end{figure}

Figure \ref{v4_v2} depicts the correlation between $v_{4}$ and $v_{2}$ over seven centrality classes from Xe-Xe collisions at $\sqrt{s_{NN}}$=5.44TeV. We have compared our model results with ALICE experimental data \cite{201882}. We observe a qualitative agreement of our model results with the experimental data. Quadrangular flow $v_{4}$ strongly increases as elliptic flow $v_{2}$ increases. This inference is supported well by the work done in article \citep{PhysRevC.92.034903}. The upper panel of the figure compares minimum bias results with the ALICE experimental data. Our HYDJET++ model predicts somewhat a non-linear positive correlation between $v_{4}$ and $v_{2}$ throughout centrality which is in fair agreement with ALICE experimental data. HYDJET++ successfully produces minimum bias results qualitatively but overestimates quantitatively in central collisions. Moving further, we see that $v_{4}$ decreases and then increases leading to a peak in semi-peripheral collisions. This peak is not so prominent in HYDJET++ results as compared to experiment. Further, as we move from semi-peripheral to most peripheral class of collisions experimental results appear a bit complex, showing a positive correlation followed by a fall (negative correlation) and then a sudden sharp rise (positive correlation). Thus, a bump and a dip is observed here. This bump is visible in our results while the dip is not observed here. HYDJET++ model results again produce a positive correlation between $v_{4}$ and $v_{2}$ and underestimates the experiment in this region thereby failing to explain the ALICE experimental result in such centrality region.

The lower panel of figure \ref{v4_v2} shows the correlation between $v_{4}$ and $v_{2}$ over seven classes of centrality from our HYDJET++ model in body-body and tip-tip geometrical configurations along with ALICE experimental data \cite{201882} for comparison.  Again, a complete agreement is observed between model and experiment qualitatively. In central collisions, body-body and tip-tip collision results are indistinguishable and overestimate the ALICE experimental data. As we move towards semi-peripheral collisions, the two geometrical configurations can be distinguished. Quantitatively, body-body results are higher than tip-tip results in this region. As we move from central to most peripheral class of collisions, results from body-body and tip-tip collisions fail to explain the experiment (underestimate the ALICE results). The bump is more clear in body-body results than in tip-tip collisions appearing early in tip-tip results.

\begin{figure}[htbp]
\centering
\includegraphics[scale=0.3]{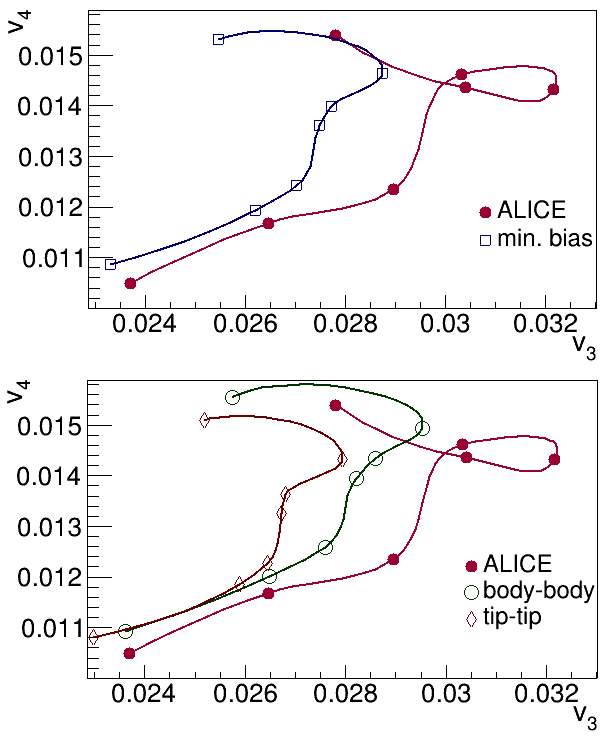} 
\caption{Correlation between $v_{4}$ and $v_{3}$ for $p_{T}<$2 GeV/c over seven centrality classes in Xe-Xe collisions at 5.44 TeV. The results have been compared with ALICE experimental data \cite{201882}.}
\label{v4_v3}
\end{figure}

Figure \ref{v4_v3} presents the correlation between $v_{4}$ and $v_{3}$ for seven centrality windows from our HYDJET++ model and from experimental data \cite{201882} for Xe-Xe collisions at the LHC energy. Here, we observe a qualitative agreement of our model results with the experimental data. The upper panel of the figure depicts the comparison of minimum bias results with the experimental data. HYDJET++ model results overpredict the data quantitatively in these centralities. In central collisions, a positive linear correlation is observed between $v_{4}$ and $v_{3}$. Moving towards higher centralities, we observe the correlation between $v_{4}$ and $v_{3}$ to be a boomerang like. Such inference is in strong agreement with the ALICE experimental data. However, model results strongly deviate from the experiment quantitatively. This is because of the fact that HYDJET++ model for the triangular flow underestimates experimental results in this region or fails to handle such collision centralities. Similar behaviour is observed for Xe-Xe collisions in body-body and tip-tip geometrical configuration, shown in lower panel of figure \ref{v4_v3}. These results have been compared with the ALICE experimental data \cite{201882} where a complete agreement is observed between model and experiment qualitatively. Quantitatively, model results overestimate the experimental data. Body-body results are higher than tip-tip results except in central collisions where the two overlap and thus it is not possible to disentangle body-body configuration from tip-tip. In central collisions, the correlation between $v_{4}$ and $v_{3}$ in both the cases is similar to the above described for the upper panel results. The HYDJET++ model results show a qualitative deviation from experimental results because the triangular flow $v_{3}$ underestimates the experimental data quantitatively in the region between mid-central to peripheral collisions.

\section{Summary and Outlook}
\label{summary}
In a brief summary of our work, we have made a scrupulous study of azimuthal anisotropic fourier harmonic coefficients in xenon-xenon collision systems at $\sqrt{s_{NN}}$= 5.44-TeV LHC energies performed under the framework of the modified HYDJET++ model, providing the possibility to study the collisions in various geometrical configurations which being cognizant of the initial conditions. Here, we have used tip-tip and body-body configurations for our analysis considering only those events which fall in the kinematic range $|\eta|<0.8$ and $0<p_{T}<2$ GeV/c and the results have been compared with ALICE experimental data. Our model results show a suitable match with the ALICE experimental data both quantitatively and qualitatively thereby enlighting both geometrical and dynamical anisotropies of the system, respectively.

We observe a strong centrality dependence of the azimuthal anisotropic harmonic coefficients $v_{2}$ and $v_{3}$ but a weak dependence of $v_{4}$ on collision centrality. In a recent article \cite{PhysRevC.103.054902}, O-O, Al-Al, and Cu-Cu collisions at 200 GeV/c as a function of centrality were studied where we observed that $v_{2}$ has a weak centrality dependence whereas both $v_{3}$ and $v_{4}$ fall and rise with centrality. Also, the quadrangular flow $v_{4}$ is observed to be larger than the triangular flow $v_{3}$, as a function of centrality. But our results are contrary to these where $v_{2}>v_{3}>v_{4}$ as a function of centrality. The minimum bias HYDJET++ model results are consistent with the ALICE experimental data. Elliptic flow results underpredict the experimental data in most peripheral class of collisions whereas triangular flow  underpredicts the experimental results except in most central and most peripheral class of collisions. However, quadrangular flow results for the HYDJET++ model are consistent with the data throughout collision centrality. Our model results in body-body and tip-tip collisions show strong dependence on collision centrality. Body-body collision results are higher than tip-tip collision results. It is possible to disentangle the two geometrical configurations in various classes of collisions. For elliptic flow, the two geometrical configuarations can be observed deviating from each other as we move from most central to most peripheral class of collisions. In case of triangular flow, the two geometrical configurations can be disentangled in all classes of collisions except in most central and most peripheral collisions. Similar is the case of quadrangular flow where body-body and tip-tip collision results overlap in most central (0\%-5\%) and in most peripheral (50\%-60\%) class of collisions.

Anisotropic flow of identified charged particles with respect to transverse momentum is studied. Again centrality dependence of elliptic, triangular and quadrangular flows is observed. Mass ordering is observed in case $v_{2},v_{3}$ and $v_{4}$. At low $p_{T}$, ($p_{T}<p_{T}^{cut}$ GeV/c) $v_{n}$ for the lower mass particle (pions) is more than the higher mass particles. However, for $p_{T}>p_{T}^{cut}$, as we move from most central to peripheral collisions higher mass particles are produced more. In most peripheral collisions, the situation is different. At $p_{T}>p_{T}^{cut}$ $v_{2}^{K}>v_{2}^{\Pi}>v_{2}^{p}$ where $m^{\Pi}<m^{K}<m^{p}$. The $p_{T}^{cut}$ values are different for $v_{2},v_{3}$ and $v_{4}$ being 1.6$\pm0.2$ GeV/c, 1.5$\pm0.4$ GeV/c, and 1.7$\pm0.2$ GeV/c, respectively. At much higher transverse momenta, $p_{T}>2.8\pm0.2$, $v_{2}^{p}$ is quite higher, the order of flow being $v_{2}^{p}>v_{2}^{K}>v_{2}^{\Pi}$.

The variation of minimum bias $v_{n}$ with respect to the total charged particle multiplicity is presented. Qualitatively, the behaviour for different flow coefficients, $v_{2},v_{3}$ and $v_{4}$ is similar to ALICE experimental results, showing a strong dependence on the total charged particle multiplicity. Quantitatively, the model almost underpredicts the experimental data.

The correlation between the different azimuthal anisotropic coefficients is also studied. Qualitative agreement of our model results with the experimental data is observed. Positive linear correlation is observed between $v_{3}$ and $v_{2}$ in central collisions. However, in mid-central or semi-peripheral collisions, correlation between $v_{3}$ and $v_{2}$ is not very much positive whereas in most peripheral collisions, a sharp negative correlation is observed having good agreement with the ALICE experimental results qualitatively. Results for both body-body and tip-tip collisions has also been presented. Quantitatively, body-body results are higher than tip-tip results. The correlation between $v_{3}$ and $v_{2}$ in both the geometrical configurations is similar to the  experimental results in central collisions. In central collisions, the two geometrical configurations are indistinguishable whereas as we move towards peripheral collisions it is very much possible to disentangle body-body collisions from tip-tip collisions.

A qualitative agreement of HYDJET++ model results with the ALICE experimental data is observed for the correlation between $v_{4}$ and $v_{2}$. Positive linear correlation between $v_{4}$ and $v_{2}$ is seen throughout centrality. HYDJET++ results overestimate experimental measurements quantitatively as we move from most central to semi-peripheral collisions. As we move from semi-peripheral to most peripheral collisions ALICE experimental results show a positive correlation followed by a fall (negative correlation) and then a sudden sharp rise (positive correlation). However, our model results show a positive correlation between $v_{4}$ and $v_{2}$ and underestimate the experimental results. Hence, failing to explain such ALICE experimental result in this region. In case of the two geometrical configurations, a suitable agreement is observed between model and experiment qualitatively. Quantitatively, body-body results are higher than tip-tip results in semi-peripheral collision region where the two configurations can be distinguished. As we move towards most peripheral collisions, body-body results being higher than tip-tip collision fail to explain the experimental data (underpredict the data).

The correlation between $v_{4}$ and $v_{3}$ for Xe-Xe collisions show a qualitative agreement of our model results with the ALICE experimental data. A positive linear correlation is observed between $v_{4}$ and $v_{3}$ in central collisions. As we move towards higher centralities, the correlation between $v_{4}$ and $v_{3}$ is observed to be a boomerang like. This is in strong agreement with the ALICE experimental data. Quantitatively, HYDJET++ model results strongly deviate from the experiment. Similar behaviour is observed in case of body-body and tip-tip collisions. Body-body collision results are higher than tip-tip results. The HYDJET++ model results show a qualitative deviation from ALICE experiment due to the reason that triangular flow $v_{3}$ underestimates the experimental data quantitatively in the region between mid-central to peripheral collisions. the two geometrical configurations are inseperable in central collisions but can be easily differentiated as we move towards most peripheral collisions.

Thus, we disintegratd our $v_{n}$-$v_{m}$ correlations into linear and non-linear contributions having strong dependence on centrality and showing strong agreement with ALICE experiment qualitatively and in some regions quantitatively. Also, at such stages we are quite successful in disentangling the geometrical configurations. Our analysis also showers some light on the geometrical and dynamical anisotropies of the system. This non-linear correlation contribution between the anisotropic coefficients may be attributed to the elliptic geometric deformation of the nuclear overlap region in non-central Xe-Xe collision systems and is visualized here since the nuclear overlap region is elliptically deformed even in most central colisions (at b=0). Further higher harmonic coefficients ($n\geq5$) can also be studied under HYDJET++ framework in Xe-Xe collision systems but due to lack of experimental evidences, the problem is a bit tacky. The study can be performed in a way by modifying the model for such higher azimuthal anisotropic harmonics and then comparing the results with inferences predicted from various thermodynamical models \cite{YAN201582, PhysRevC.94.024910}. We leave this part for our future work.

\section*{ACKNOWLEDGEMENTS}
\vspace{-2ex}
We sincerely acknowledge financial support from the Institutions of Eminence (IoE) BHU grant. SP acknowledges the financial support obtained from UGC under research fellowship scheme during the work.
\bibliographystyle{apsrev}

\bibliography{references}

\end{document}